# Investigation of self-focusing of Gaussian laser beams within magnetized plasma via source-dependent expansion method

A. A. Molavi Choobini[1], S. S. Ghaffari-Oskooei[2*]

[1]Dept. of Physics, University of Tehran, Tehran 14399-55961, Iran,

[2]Department of Atomic and Molecular Physics, Faculty of Physics, Alzahra University, Tehran, Iran.

**Abstract:**

Self-focusing emerges as a nonlinear optical phenomenon resulting from an intense laser field and plasma interaction. This study investigates the self-focusing behavior of Gaussian laser beams within magnetized plasma environments utilizing a novel approach, source-dependent expansion. By employing source-dependent expansion, we explore the intricate dynamics of laser beam propagation, considering the influence of plasma density and external magnetic fields. The interplay between the beam's Gaussian profile and the self-focusing mechanism through rigorous mathematical analysis and numerical simulations, particularly in the presence of plasma-induced nonlinearities are elucidated here. Our findings reveal crucial insights into the evolution of laser beams under diverse parameters, the ponderomotive force, relativistic factors, plasma frequency, polarization states, external magnetic field, wavelength, and laser intensity. This research not only contributes to advancing our fundamental understanding of laser-plasma interactions but also holds promise for optimizing laser-driven applications.

## I. Introduction

In the dynamic realm of scientific inquiry, the phenomenon of laser-plasma interaction has become an enthralling subject for both experimental and theoretical researchers over the past two decades, primarily driven by its diverse applications spanning particle acceleration, and the generation of high-frequency radiation sources [1, 2]. The multifaceted nature of these applications has spurred a comprehensive exploration of the intricate dynamics underlying laser-plasma interaction. In the majority of the scenarios mentioned, the essential requirement is for a laser pulse to extend its propagation across several Rayleigh lengths [3, 4]. However, the achievement of this objective is hindered by the inherent tendency of the laser pulse to undergo defocusing. The complexities of experimental studies on laser-plasma interaction are compounded by various effects stemming from the non-linear nature of laser-plasma dynamics. A breakthrough in this challenge arises from the ability of high-intense laser pulse to surmount natural diffractive defocusing [5]. This is accomplished by maintaining focus through the nonlinear adjustment of the dielectric constant of the plasma [6]. The laser, marked by its non-uniform intensity, applies a transverse ponderomotive force, displacing electrons from the longitudinal axis and guiding them towards areas with lower intensity [7, 8]. This intricate interplay between the laser and plasma components is pivotal in preserving the focus of the laser pulse against the tendency to defocus, enabling more effective and controlled interactions in experimental settings. The ponderomotive force changes the plasma's local effective permittivity, thereby impacting the focal properties of the laser according to the altered plasma parameters and laser power [9, 10].

M. Dwivedi and H. K. Malik explored the dynamics of low-density channels formed through the outward expulsion of electrons and ions in a radial direction under the influence of the ponderomotive force generated by a q-Gaussian laser, employing the WKB approximation [11]. G. Zhang and colleagues conducted a study on relativistic self-focusing in the interaction between a laser beam and plasma featuring periodic density ripples, utilizing a higher-order paraxial theory [12]. Their findings reveal that, influenced by relativistic nonlinear effects, the dielectric function undergoes vigorous oscillations with a corresponding periodicity. K. Walia's work delves into the self-focusing behavior of high-power beams in unmagnetized plasma and its impact on the Stimulated Raman Scattering (SRS) process, considering both relativistic-ponderomotive nonlinearities [13]. His investigation affirms that the focusing of the involved beams results in an elevation of SRS back-reflectivity. V. S. Pawar et al. conducted a study on the phenomenon of relativistic self-focusing which involved finite Airy-Gaussian laser beams within a cold quantum plasma, employing standard WKB and paraxial approximations [14]. Their research investigates the influence of the modulation parameter, critical intensity parameter, and critical initial beam radius on the self-focusing phenomenon. M. Martyanov and his colleagues conducted an assessment of the suppression of small-scale self-focusing in high-power femtosecond laser pulses [15]. Experimental findings demonstrated that the impact of filamentation instability, often referred to as small-scale self-focusing, is notably weaker than the forecasts from both stationary and nonstationary theoretical frameworks for high B-integral values. H. Nie and collaborators introduced an innovative strategy to counteract small-scale self-focusing (SSSF) in high-power lasers (HPLs) [16]. Their approach involves employing rotating beams created through the coherent superposition of two vortex beams with opposite topological charges and frequency shifts. This method effectively reduces the breakup integral and mid-high frequency components of the HPLs, offering a novel means to mitigate the challenges associated with small-scale self-focusing. A. A. Butt et al., have examined the combined effect of a wiggler magnetic field and an exponential plasma density ramp on laser beam of q-Gaussian type in under-dense plasma [17]. Their observations disclosed that the incorporation of these parameters resulted in a decrease in the beam width parameter of the q-GLB. This, in turn, contributed to an intensified self-focusing effect, achieved by displacing plasma electrons away from the central axis of the laser beam. In addition, the previous research has delved into Raman and Brillouin scattering from self-focused Gaussian laser beams in plasma which the intensity of scattered waves is derived usage of the paraxial approximation [18].

In contrast to the aforementioned studies, our current investigation employs a technique known as source-dependent expansion within a magnetized plasma, incorporating both ponderomotive force and relativistic effects. This approach demonstrates superior applicability across a wider spectrum of magnetized plasma parameters and geometries compared to the WKB approximation.

It excels in scenarios where wave frequencies are comparable to or even smaller than the characteristic frequencies of the plasma, where the WKB approximation falters. Moreover, source-dependent expansion facilitates eigenmode analysis, enabling the determination of mode structures and eigenfrequencies in magnetized plasmas. Thus, when heightened accuracy and broader applicability are imperative, source-dependent expansion emerges as the preferred alternative to the WKB approximation. The paper is arranged as follows: in section II, the envelope equation, which refers to the description of how the spot size changes with the normalized distance of propagation, is derived through source-dependent expansion. Relativistic effects along with ponderomotive forces are considered in the theoretical model. Section III presents the numerical solution of the spot size equation where the effect of laser and plasma parameters on the evolution of laser spot size and intensity is investigated. Conclusions are drawn in section IV.

## II. Analytical Model

Consider a circularly-polarized laser pulse which propagates in a magnetized plasma. Plasma is presumed to be inserted in a constant magnetic field $\vec{B}_0 = B_0 \hat{z}$. In this context, the vector potential of the laser pulse conforms to Maxwell's wave equation:

$$-\nabla^2 \vec{A} + \frac{1}{c^2} \frac{\partial^2 \vec{A}}{\partial t^2} = \frac{4\pi}{c} \vec{J} \quad (1)$$

where $\vec{A}$ represents the vector potential of laser pulse:

$$\vec{A}(r,z,t) = \frac{1}{2} A(r,z)(\hat{x} + i\sigma\hat{y})e^{ikz-i\omega t} + c.c. \quad (2)$$

here, $k$ and $\omega$ are wave-vector and frequency of laser, respectively. Parameter of $\sigma = \pm 1$ indicates the right and left-handed polarizations, and $\vec{J}$ is the current density of electrons as follows:

$$\vec{J} = -(n_0 + \delta n)e\vec{v} \quad (3)$$

where $n_0$ and $\delta n$ are ambient and perturbated densities of plasma. By defining normalized vector potential, $\vec{a} = -\frac{e\vec{A}}{mc^2}$ where $\vec{a}(r,z,t) = \frac{1}{2}a(r,z)(\hat{x} + i\sigma\hat{y})e^{ikz-i\omega t} + c.c.$, and substitution of electrons velocity $\vec{v} = -\frac{c\vec{a}}{\gamma(1-\frac{\sigma\omega_c}{\omega})}$ in Eq. (1), where the relativistic factor of $\gamma$ is represented by $\gamma = \sqrt{1 + |a|^2 / \left(1 - \frac{\sigma\omega_c}{\omega}\right)^2}$. Therefore, one can derive the following equation in paraxial approximation:

$$\left(-\frac{\omega^2}{c^2} + k^2 - \nabla_\perp^2 - 2ik\frac{\partial}{\partial z}\right)a = -\frac{\omega_p^2}{\gamma c^2}\left(1 + \frac{\delta n}{n_0}\right)\frac{1}{\left(1-\frac{\sigma\omega_c}{\gamma\omega}\right)}a \quad (4)$$

Equalization of ponderomotive and space charge forces leads to the following equation for $\frac{\delta n}{n_0}$:

$$\frac{\delta n}{n_0} = \frac{c^2}{\omega_p^2}\nabla_\perp^2 \gamma \quad (5)$$

Therefore, Eq. (4) can be re-written as:

$$\left(\nabla_\perp^2 + 2ik\frac{\partial}{\partial z}\right)a = k^2(1-\eta^2)a \tag{6}$$

where

$$\eta^2 = \frac{\omega^2}{c^2 k^2} - \frac{\omega_p^2}{c^2 k^2 \gamma \left(1-\frac{\sigma\omega_c}{\gamma\omega}\right)}\left(1 + \frac{c^2}{\omega_p^2}\nabla_\perp^2 \gamma\right) \tag{7}$$

To solve Eq. (6), the amplitude of $a(r,z)$ can be expanded by source-dependent expansion (SDE) as follows:

$$a(r,z) = \sum_{m=0}^{\infty} a_m(z)D_m(r,z) \tag{8}$$

where

$$D_m(r,z) = L_m\left(\frac{2r^2}{r_{s(z)}^2}\right)\exp\left(-(1-i\alpha_s)\frac{r^2}{r_{s(z)}^2}\right) \tag{9}$$

Here, $L_m$ is the Laguerre polynomial of order m. The quantities of $\alpha_s$ and $r_s$ represent the curvature of wavefront and spot size of laser pulse, respectively. By insertion of Eq. (8) in Eq. (6) and assuming that $|a_0| \gg |a_{m>0}|$, one can derive the following partial differential equation for spot size of laser pulse:

$$\frac{\partial^2 r_s}{\partial z^2} - \frac{4}{k^2 r_s^3} - \frac{4}{r_s}G = 0 \tag{10}$$

here with definition of $\chi = \frac{2r^2}{r_{s(z)}^2}$, the function of G has an integral form as follows:

$$G = \frac{1}{2}\int_0^\infty d\chi(1-\eta^2)(1-\chi)\exp(-\chi) \tag{11}$$

where G can be written as the summation of three integrals for relativistic, magnetized and ponderomotive effects where each integral shows the contribution a specific factor in self-focusing:

$$G = G_{relativistic} + G_{magnetized} + G_{ponderomotive} \tag{12}$$

where

$$G_{rel} = \frac{1}{2}\left(\frac{\omega_p^2}{c^2 k^2}\right)\int_0^\infty \frac{(1-\chi)\exp(-\chi)}{\sqrt{1+\frac{a_0^2 \exp(-\chi)}{\left(1-\frac{\sigma\omega_c}{\omega}\right)^2}}} d\chi \tag{13}$$

$$G_{mag} = \frac{1}{2}\left(\frac{\sigma\omega_c}{\omega}\right)\int_0^\infty \frac{(1-\chi)\exp(-\chi)}{\left(1+\frac{a_0^2 \exp(-\chi)}{\left(1-\frac{\sigma\omega_c}{\omega}\right)^2}\right)}\left(\frac{\omega_p^2}{c^2 k^2} + \frac{\nabla_\perp^2 \gamma}{k^2}\right) d\chi \tag{14}$$

$$G_{pond} = \left(\frac{1}{2k^2}\right) \int_0^\infty \frac{(1-\chi)\exp(-\chi)\nabla_\perp^2 \gamma}{\sqrt{1+\frac{a_0^2 \exp(-\chi)}{\left(1-\frac{\sigma\omega_c}{\omega}\right)^2}}} d\chi \tag{15}$$

The integrals associated with ponderomotive effects typically involve quantities related to the ponderomotive potential, relativistic and magnetized effects which describe the energy exchange between the electromagnetic field and the plasma particles. The ponderomotive force can cause density modulations in the plasma, leading to self-focusing or defocusing depending on the balance of forces involved. The magnetic field can confine the plasma and influence the self-focusing process by altering the dynamics of the plasma and the electromagnetic fields involved. For self-focusing, relativistic effects can enhance the focusing mechanism by increasing the effective mass of charged particles, which affects the plasma's response to the electromagnetic fields driving the self-focusing process. Therefore, three integrals for G can be evaluated through the following relations:

$$G_{rel} = \frac{1}{6a_0^2}\left(\frac{\omega_p^2}{c^2 k^2}\right)\left(1-\frac{\sigma\omega_c}{\omega}\right)^2 \left[6\ln\left(1+\frac{a_0^2}{2\left(1-\frac{\sigma\omega_c}{\omega}\right)^2}\right) - 3\ln^2\left(\frac{a_0^2}{2\left(1-\frac{\sigma\omega_c}{\omega}\right)^2}\right) - 6Li_2\left(-\frac{2\left(1-\frac{\sigma\omega_c}{\omega}\right)^2}{a_0^2}\right) - \pi^2\right] \tag{16}$$

$$G_{mag} = \frac{1}{12a_0^2}\left(\frac{\omega_p^2}{c^2 k^2}\right)\left(\frac{\sigma\omega_c}{\omega}\right)\left(1-\frac{\sigma\omega_c}{\omega}\right)^2 \left[6\ln\left(1+\frac{a_0^2}{\left(1-\frac{\sigma\omega_c}{\omega}\right)^2}\right) - 3\ln^2\left(\frac{a_0^2}{\left(1-\frac{\sigma\omega_c}{\omega}\right)^2}\right) - 6Li_2\left(-\frac{\left(1-\frac{\sigma\omega_c}{\omega}\right)^2}{a_0^2}\right) - \pi^2\right] + \left(\frac{-1}{12k^2 a_0^2 r_s^2}\right)\left(\frac{\sigma\omega_c}{\omega}\right)\left(1-\frac{\sigma\omega_c}{\omega}\right)^2 \left[12\ln\left(1+\frac{a_0^2}{2\left(1-\frac{\sigma\omega_c}{\omega}\right)^2}\right) - 6\ln^2\left(\frac{a_0^2}{2\left(1-\frac{\sigma\omega_c}{\omega}\right)^2}\right) + 12\ln(2)\ln\left(\frac{a_0^2}{2\left(1-\frac{\sigma\omega_c}{\omega}\right)^2}\right) - 12Li_2\left(-\frac{2\left(1-\frac{\sigma\omega_c}{\omega}\right)^2}{a_0^2}\right) - 6\ln^2(2) - 2\pi^2\right] \tag{17}$$

$$G_{pond} = \left(\frac{-1}{6k^2 a_0^2 r_s^2}\right)\left(1-\frac{\sigma\omega_c}{\omega}\right)^2 \left[12\ln\left(1+\frac{a_0^2}{2\left(1-\frac{\sigma\omega_c}{\omega}\right)^2}\right) - 6\ln^2\left(\frac{a_0^2}{2\left(1-\frac{\sigma\omega_c}{\omega}\right)^2}\right) + 12\ln(2)\ln\left(\frac{a_0^2}{2\left(1-\frac{\sigma\omega_c}{\omega}\right)^2}\right) - 12Li_2\left(-\frac{2\left(1-\frac{\sigma\omega_c}{\omega}\right)^2}{a_0^2}\right) - 6\ln^2(2) - 2\pi^2\right] \tag{18}$$

where $Li_2$ is the dilogarithm function [19]. Self-focusing of laser pulses in plasmas is investigated in the next section by numerical solving of Eq. 10.

## III. Results and Discussion

Self-focusing is a nonlinear optical phenomenon that arises due to the interaction of the intense laser field and the electrons of plasma. The study of the self-focusing of a laser beam within a magnetized plasma is essential for optimizing laser-plasma interactions in various applications and exploring new avenues in high-energy physics and optical engineering. Laser-driven plasma devices are used in various applications such as particle accelerators, plasma-based particle acceleration and radiation sources, and plasma lenses for focusing charged particle beams. Understanding self-focusing helps in designing and optimizing the performance of these devices.

Self-focusing enhances the intensity and confinement of the laser beam, leading to more efficient particle acceleration and higher energy radiation sources (Fig. 1). The experimental parameters listed in references [20, 21] were considered for simulation runs.

Figure 2 demonstrates the impact of the ponderomotive force on the amplitude of the spot size of the laser beam. It illustrates that when relativistic effects are incorporated into the ponderomotive force, there is an observed increase in the amplitude of the spot size of the laser beam, accompanied by self-focusing. The effect of the ponderomotive force on the spot size of a laser beam is related to the self-focusing phenomenon. Self-focusing occurs when the intensity of the laser beam is high enough to induce a nonlinear response in the medium through which it propagates. When the ponderomotive force is present, it can cause a modification of the refractive index of the medium. This modification is typically nonlinear, meaning it depends on the intensity of the laser beam. As a result, the refractive index of the medium becomes intensity-dependent, leading to a change in the behavior of the laser beam. In the case where the relativistic effects are added to the ponderomotive force, the amplitude of the spot size of the laser beam tends to increase, and self-focusing occurs. The relativistic effects can enhance the strength of the ponderomotive force, which in turn increases the intensity-dependent refractive index modification. This enhanced modification causes the laser beam to experience a focusing effect, leading to a decrease in the spot size. Furthermore, coupling the relativistic effects with the ponderomotive force intensifies the interaction of the laser beam with the charged particles, ultimately causing the observed increase in spot size amplitude and self-focusing phenomena. The ponderomotive force is augmented by relativistic effects, which account for the high velocities and energies of particles approaching the speed of light, it results in a more pronounced influence on the laser beam. Specifically, the increased amplitude of the spot size suggests a greater concentration or focusing of the laser beam's energy, leading to self-focusing behavior.

The variations in spot size of the laser beam versus length for different relativistic factors are depicted in Figure 3. According to the figure, as the relativistic factor is enhanced, there is a noticeable increase in the amplitude of the spot size of the laser beam, although the enhancement is not significant. Additionally, it is observed that the period and wavelength of self-focusing of the beam remain unchanged. This phenomenon can be attributed to the interplay between the relativistic effects and the characteristics of the laser beam. As the relativistic factor is increased, the influence of relativistic effects on the ponderomotive force becomes more pronounced, leading to a slight boost in the amplitude of the spot size. However, the period and wavelength of self-focusing remain unaffected because these parameters are primarily determined by the intrinsic properties of the laser beam and the medium through which it propagates, rather than solely by the relativistic factor. On the other hand, the behavior of self-focusing and spot size amplification can be subject to saturation effects. As the relativistic factor increases, the amplification of the spot size may reach a saturation point beyond which further increases in the relativistic factor have diminishing effects. Furthermore, other nonlinear effects may dominate the self-focusing behavior, overshadowing the influence of the relativistic factor on the period and wavelength of self-focusing. Nonlinear effects such as Kerr nonlinearity or plasma effects can play a significant role in determining the period and wavelength of self-focusing. If these effects are more pronounced or have a stronger influence than the relativistic factor, their impact on the self-focusing behavior may overshadow any variations caused by the relativistic factor.

Figure 4 displays the effect of plasma frequency on the variation of spot size of a laser beam. This figure indicates that as the plasma frequency or ambient density of plasma increases, the spot size of the beam is enhanced, and the number of peaks is decreased. As the plasma frequency increases, the refractive index of the plasma increases, causing greater refraction of the laser beam. This refraction can cause the beam to spread out more, leading to an increase in spot size. As the plasma frequency or ambient plasma density increases, the intensity-dependent refractive index modification becomes more significant. This enhanced modification causes a stronger defocusing effect on the laser beam, resulting in an increased spot size. The defocusing effect counteracts the natural tendency of the laser beam to focus due to diffraction, causing the beam to spread out and have a larger spot size. At certain plasma densities and frequencies, the plasma can exhibit self-focusing or self-defocusing effects on the laser beam. When the plasma frequency is low, self-focusing dominates, causing the laser beam to concentrate, which can reduce the spot size. Conversely, at higher plasma frequencies, self-defocusing may occur, causing the beam to spread out and increasing the spot size. As the plasma frequency increases, the interaction of the laser beam with the electrons of plasma becomes stronger. This can lead to more significant energy transfer from the laser beam to the plasma, resulting in energy loss and a decrease in the intensity of the laser beam. Consequently, the reduced intensity leads to a broader spot size distribution. This spreading effect reduces the sharpness and distinctness of the peaks, leading to a decrease in the number of observable peaks.

The effect of the various polarization states on the variations in the spot size of the laser beam is investigated. The differences in spot size variation observed among different polarization states of the laser beam, as depicted in Fig. 5, can be attributed to the interaction between the polarization state of the beam and the properties of the medium through which it propagates. As shown, in the right-hand polarization, the spot size of the laser beam has maximum amplitude compared to the left-hand or linear polarization. The maximum amplitude of spot size variation observed for right-hand polarization compared to left-hand or linear polarization in a plasma environment is influenced by a combination of nonlinear optical effects, self-focusing behavior, plasma resonance and absorption, anisotropic properties, and the plasma's response to polarization. These factors collectively shape the interaction of the laser beam with the electrons of plasma, resulting in differential effects on spot size variation for different polarization states. If a laser beam propagates through a plasma medium, the presence of the plasma can induce birefringence, which is the splitting of the beam into two orthogonal polarization components. This birefringence arises due to the anisotropic response of the plasma to the electromagnetic field of the laser beam. The birefringence induced by the plasma be visualized using a refractive index ellipsoid, which represents the anisotropic nature of the plasma medium. In the case of plasma-induced birefringence, the refractive index ellipsoid becomes elongated along one axis and compressed along another axis. Due to the plasma-induced birefringence, the laser beam experiences different focusing characteristics for different polarization states. In the case of Figure 5, the right-hand polarization state aligns with the elongated axis of the refractive index ellipsoid, which results in a weaker focusing effect and a larger spot size compared to the left-hand or linear polarization states. In addition, plasma has resonance frequencies or absorption characteristics that favor certain polarization states over others. In this specific case, the right-hand polarization coincides with resonance frequencies and experiences enhanced absorption within the plasma, leading to stronger interactions and more significant variations in spot size compared to left-hand or linear polarization.

In Figure 6, the role of the external magnetic field on the variation of spot size of a laser beam is showcased across other distinct sets of parameters. The analysis of the figure indicates that an increase in the external magnetic field leads to an increase in the amplitude of the spot size. However, the number of periodic is boosted. This is due to the fact that when a laser beam propagates through a plasma in the presence of an external magnetic field, the interaction between the magnetic field and the plasma can give rise to magnetic self-focusing. Magnetic self-focusing occurs when the magnetic field causes the plasma to become a focusing medium for the laser beam. The external magnetic field exerts a force on the plasma, compressing it in the transverse direction perpendicular to the magnetic field lines. This compression leads to an increase in plasma density in those regions. The increased plasma density, in turn, results in a higher refractive index for the laser beam in those regions. The increased plasma density and the corresponding increase in refractive index due to the external magnetic field create a self-focusing effect on the laser beam. This means that the laser beam experiences enhanced focusing as it propagates through the plasma. As the external magnetic field increases, the self-focusing effect becomes stronger. The enhanced self-focusing leads to a more pronounced reduction in the spot size of the laser beam, increasing the amplitude of the spot size. This means that the beam becomes more tightly focused, resulting in a smaller spot-size amplitude. In contrast, the self-focusing effect in the presence of an external magnetic field can also lead to the formation of periodic structures in the laser beam, commonly referred to as filaments. These filaments are regions of higher intensity and narrower spot sizes within the laser beam. As the self-focusing becomes more pronounced with increasing magnetic field, the number of filaments or periodic within the beam increases. These filaments can act as waveguides, confining and guiding the laser beam along their lengths. The presence of plasma filaments can result in periodic modulations of the laser beam's intensity profile, leading to the observed boost in the number of periods in the spot size variation. Furthermore, the external magnetic field can modify the plasma frequency, which is the frequency at which plasma oscillates in response to electromagnetic fields. Changes in the plasma frequency can alter the interaction of the laser beam with the electrons of plasma, leading to variations in the spot size and the number of periodic observed in the spot size variation.

Figure 7 illustrates the impact of the wavelength of the laser beam on the amplitude of the spot size of the laser beam. The figure indicates that shorter wavelengths tend to exhibit smaller spot sizes compared to longer wavelengths. When a laser beam interacts with a plasma, several processes come into play, including absorption, scattering, and refraction. The behavior of these processes depends on the wavelength of the laser beam. Shorter wavelengths correspond to higher energy photons. In the context of plasma physics, shorter wavelengths can penetrate deeper into the plasma before being absorbed or scattered. This deeper penetration can result in a more localized interaction between the laser beam and the plasma, leading to a smaller spot size. Furthermore, shorter wavelengths can also undergo less scattering and diffraction within the plasma compared to longer wavelengths. This reduced scattering and diffraction contribute to maintaining the spatial coherence of the laser beam, resulting in a smaller spot size.

The variations in spot size of the laser beam versus length for different laser intensities are depicted in Fig. 8. As shown, when the intensity of the laser beam is closer to the relativistic intensities, the spot size of the laser is increased. At relativistic intensities, the electric field of the laser beam becomes so strong that it interacts significantly with the plasma medium through which it propagates. This interaction can lead to a phenomenon known as relativistic self-focusing, where the laser beam concentrates its energy into a smaller region due to the ponderomotive force exerted on the plasma electrons. As the laser intensity approaches relativistic levels, the relativistic self-focusing effect becomes more pronounced. This enhanced self-focusing leads to a reduction in the spot size of the laser beam, resulting in tighter focusing. However, at extremely high intensities, the self-focusing effect can become so strong that it leads to an increase in the spot size of the laser beam. This is because the intense laser field can induce nonlinear effects that result in the beam broadening or diffusing over the propagation distance. On the other hand, at relativistic intensities, the plasma exhibits highly nonlinear behavior. The interaction of the laser beam with the electrons of plasma can lead to the generation of plasma waves, which can affect the propagation of the laser beam. These nonlinear effects can cause the laser beam to spread out over a larger area, increasing in spot size. In addition, at extremely high intensities, photon-photon scattering becomes significant. This phenomenon involves the interaction of photons within the laser beam, leading to the creation of electron-positron pairs. As a result, the effective index of refraction of the plasma changes, which can disturb the focusing properties of the laser beam and contribute to an increase in spot size.

## IV.  Conclusions

Self-focusing of the Gaussian laser beam within a magnetized plasma is investigated through source-dependent expansion for the first time. The envelope equation, which describes the spot size of the Gaussian beam, is derived analytically where the ponderomotive force and relativistic effect are considered. The partial differential equation, which describes the evolution of the spot size of the Gaussian beam propagating in magnetized plasma, is solved numerically for different laser and plasma parameters. Plots of the spot size indicate that the spot size reaches a specific maximum and then falls to its initial value. This phenomenon, which is called self-focusing, occurs in a plasma medium if appropriate laser and plasma parameters are selected. Laser self-focusing is crucial for experimental applications of laser and plasma interactions since it facilitates the propagation of laser beams over far distances. Plots of spot size indicate that the increase of relativistic factor and plasma density as well as laser frequency and its intensity leads to the enhancement of maximum values of laser spot size which means that the spot size of the laser oscillates in a wider domain. The influence of DC magnetic field and laser polarization on the self-focusing of the laser beams in magnetized plasma is also investigated in the present study. The left and right-hand circularly polarized laser beams show different behaviors which is the result of the anisotropic properties of magnetized plasma. The increase of DC magnetic field decreases the spot size of the right-handed circularly polarization state of the laser beam which indicates that using magnetized plasma can be useful for guiding laser beams across several Rayleigh lengths.

## Acknowledgment

This research did not receive any specific grant from funding agencies in the public, commercial, or not-for-profit sectors.

**List of Figures & Captions**

**Fig. 1.** Schematic of self-focusing of laser beam in magnetized plasma.

**Fig. 2.** Impact of the ponderomotive force on the amplitude of spot size of the laser beam versus the normalized distance of propagation ($\zeta$) for $\lambda = 800nm$, $\gamma = 1.1$, $\omega_p = 2THz$, $B_0 = 100T$, $\sigma = +1$ and $a = 0.01$.

**Fig. 3.** Variations in spot size of laser beam versus the normalized distance of propagation ($\zeta$) for different relativistic factors and for $\lambda = 800nm$, $\omega_p = 2THz$, $B_0 = 100T$, $\sigma = +1$ and $a = 0.01$.

**Fig. 4.** The effect of plasma frequency on variation of spot size of a laser beam versus the normalized distance of propagation ($\zeta$) for $\lambda = 800nm$, $\gamma = 1.1$, $B_0 = 100T$, $\sigma = +1$ and $a = 0.01$.

**Fig. 5.** Variations of spot size of laser beam versus the normalized distance of propagation ($\zeta$) for various polarization states and for $\lambda = 800nm$, $\gamma = 1.1$, $\omega_p = 2THz$, $B_0 = 100T$ and $a = 0.01$.

**Fig. 6.** The role of the external magnetic field on variation of spot size of a laser beam versus the normalized distance of propagation ($\zeta$) for $\lambda = 800nm$, $\gamma = 1.1$, $\omega_p = 2THz$, $\sigma = +1$ and $a = 0.01$.

**Fig. 7.** Impact of the wavelength of the laser beam on the amplitude of spot size of the laser beam versus the normalized distance of propagation ($\zeta$) for $\gamma = 1.1$, $\omega_p = 2THz$, $B_0 = 100T$, $\sigma = +1$ and $a = 0.01$.

**Fig. 8.** Variations in spot size of laser beam versus the normalized distance of propagation ($\zeta$) for different laser intensities and for $\lambda = 800nm$, $\gamma = 1.1$, $\omega_p = 2THz$, $B_0 = 100T$ and $\sigma = +1$.

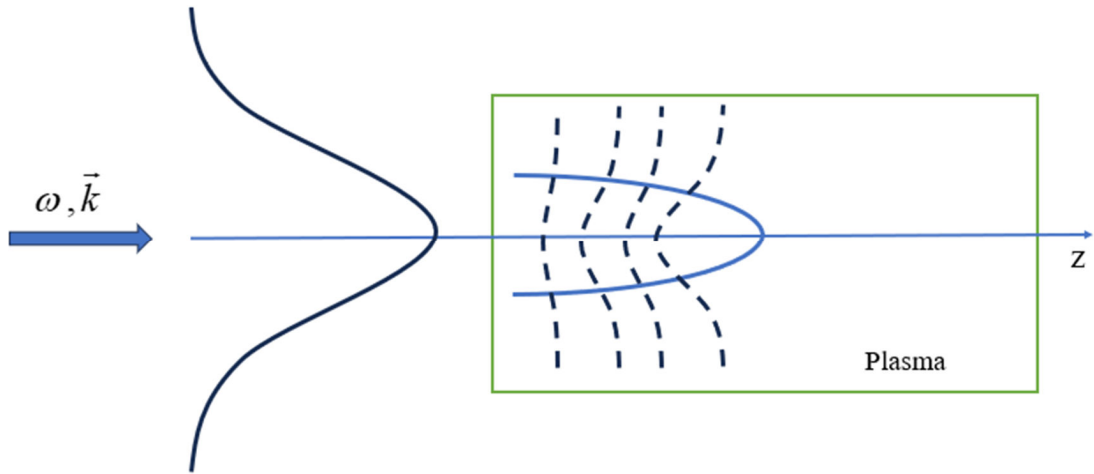

**Fig. 1.**

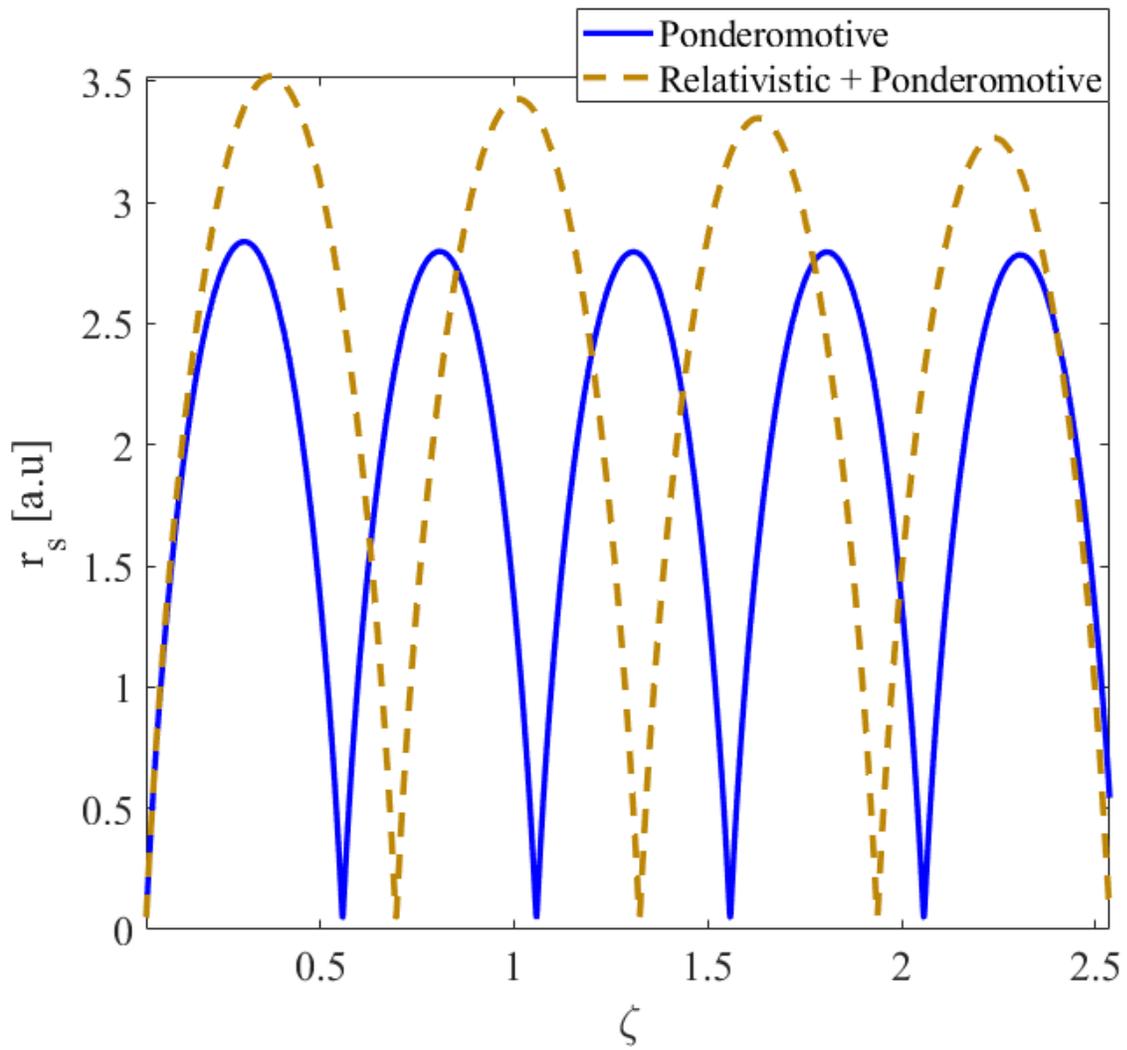

**Fig. 2.**

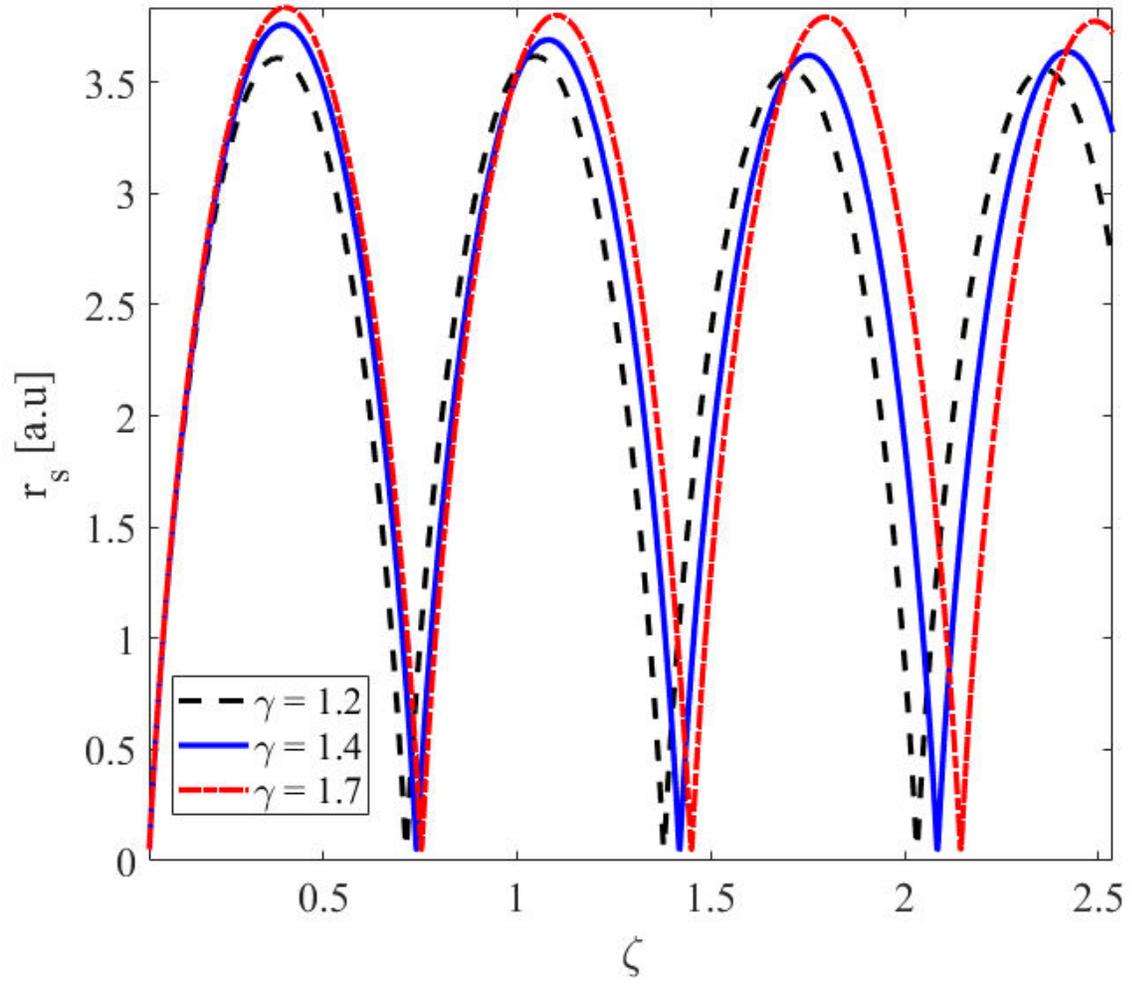

**Fig. 3.**

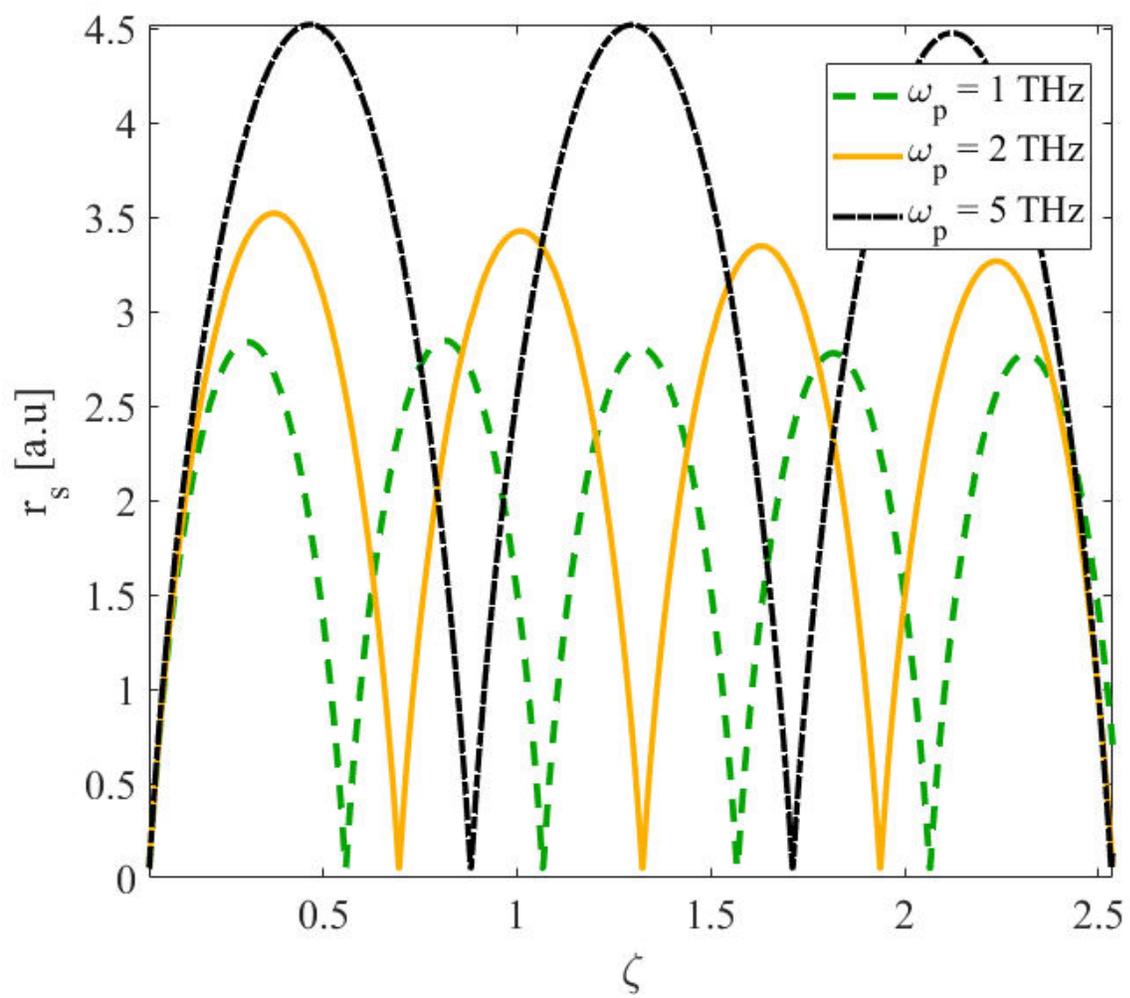

**Fig. 4.**

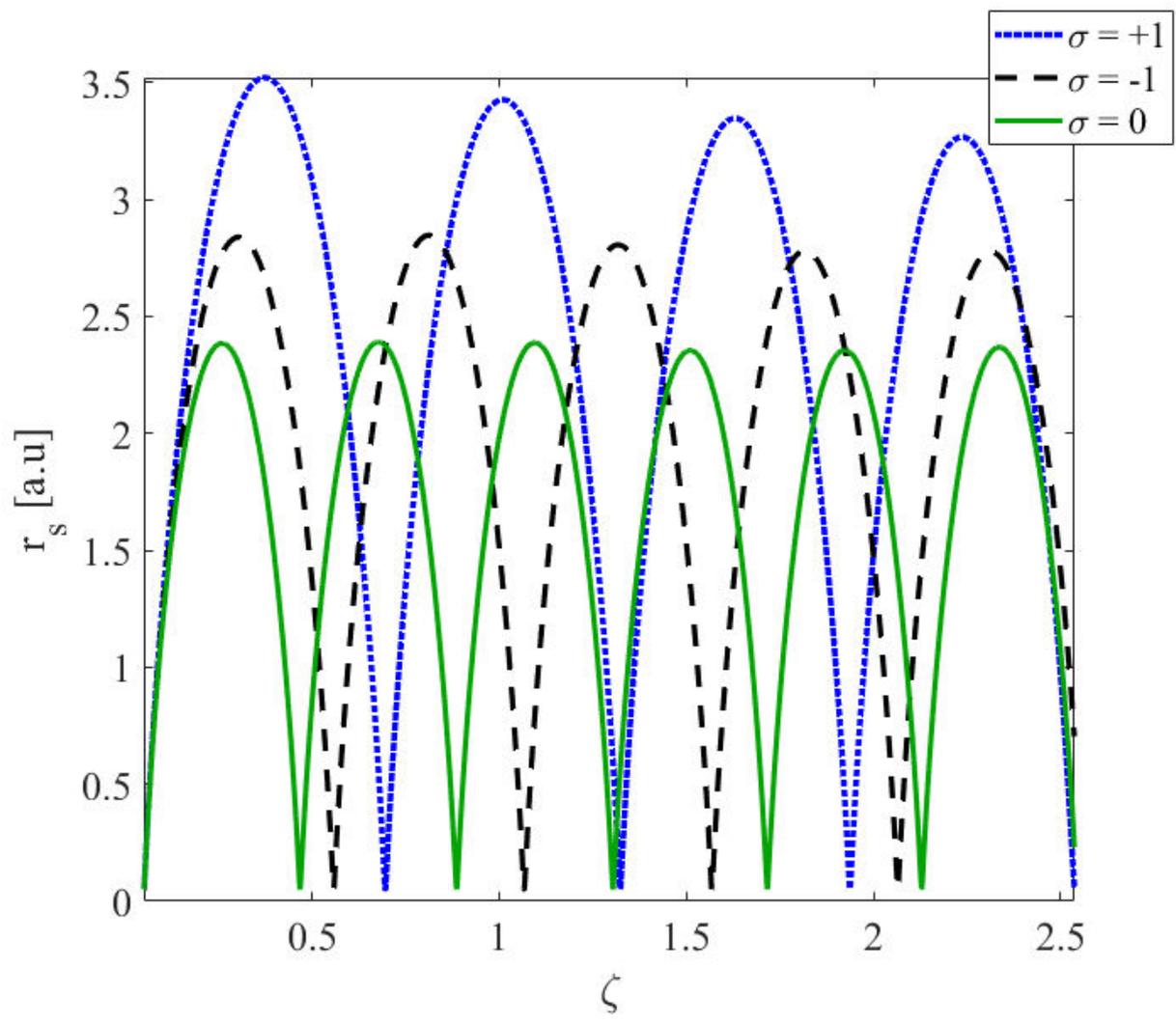

**Fig. 5.**

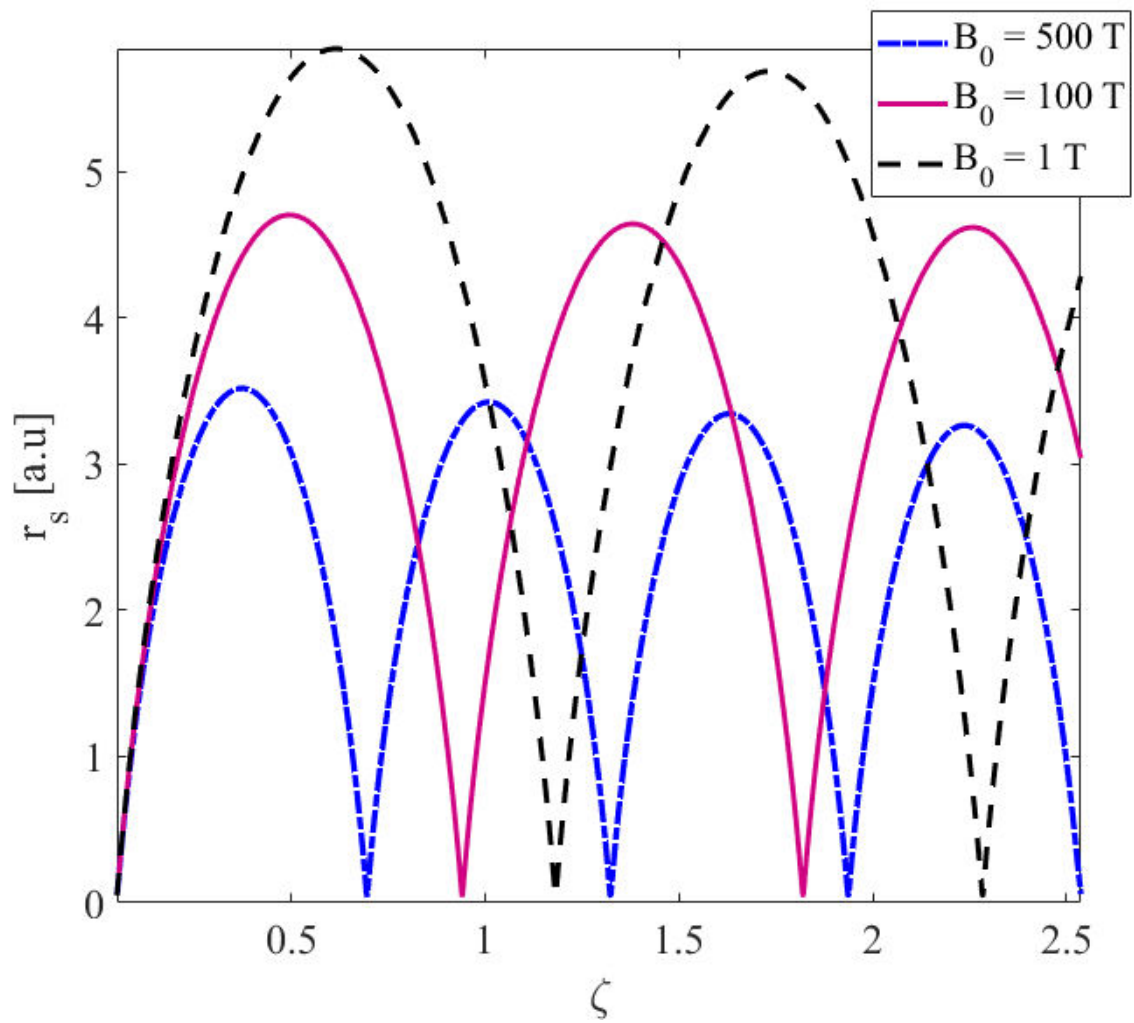

**Fig. 6.**

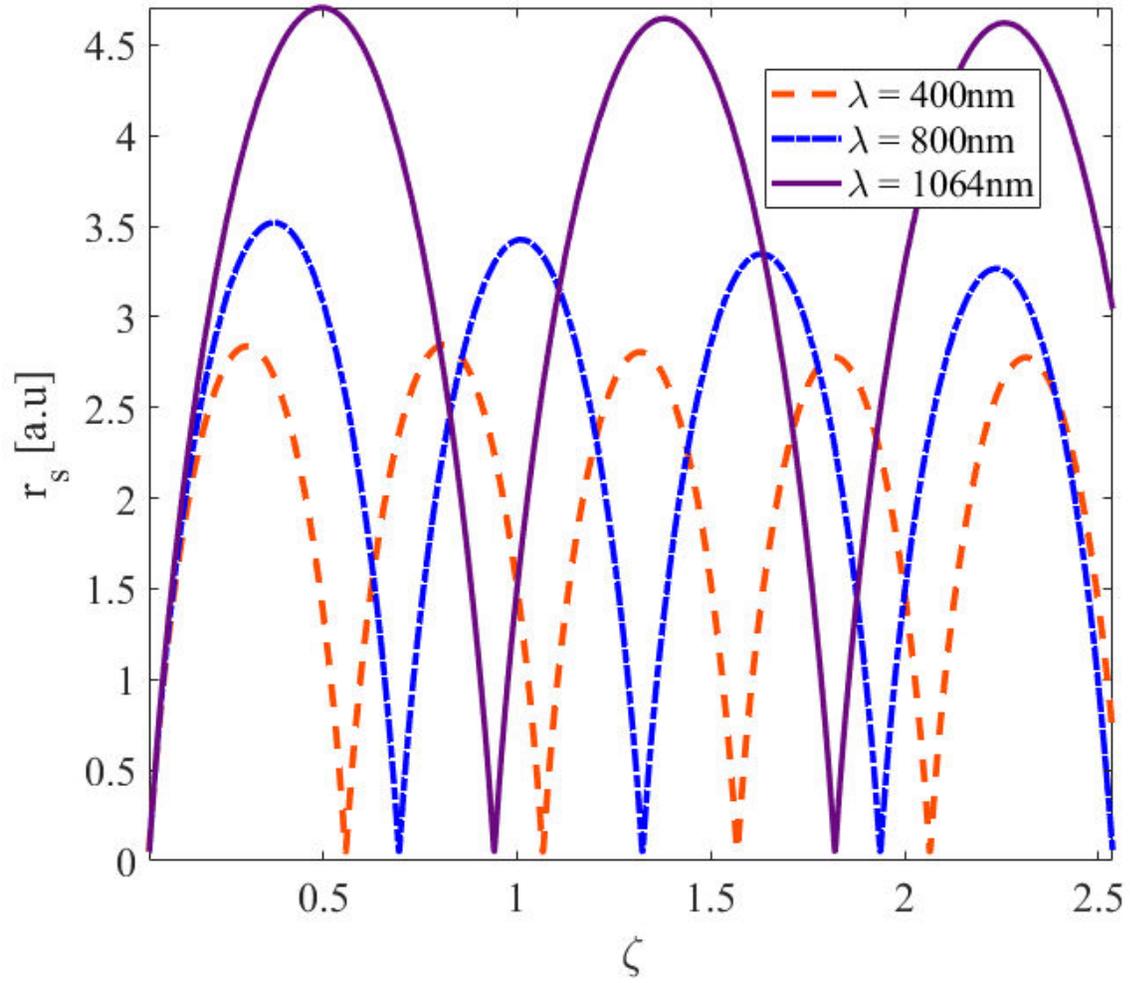

**Fig. 7.**

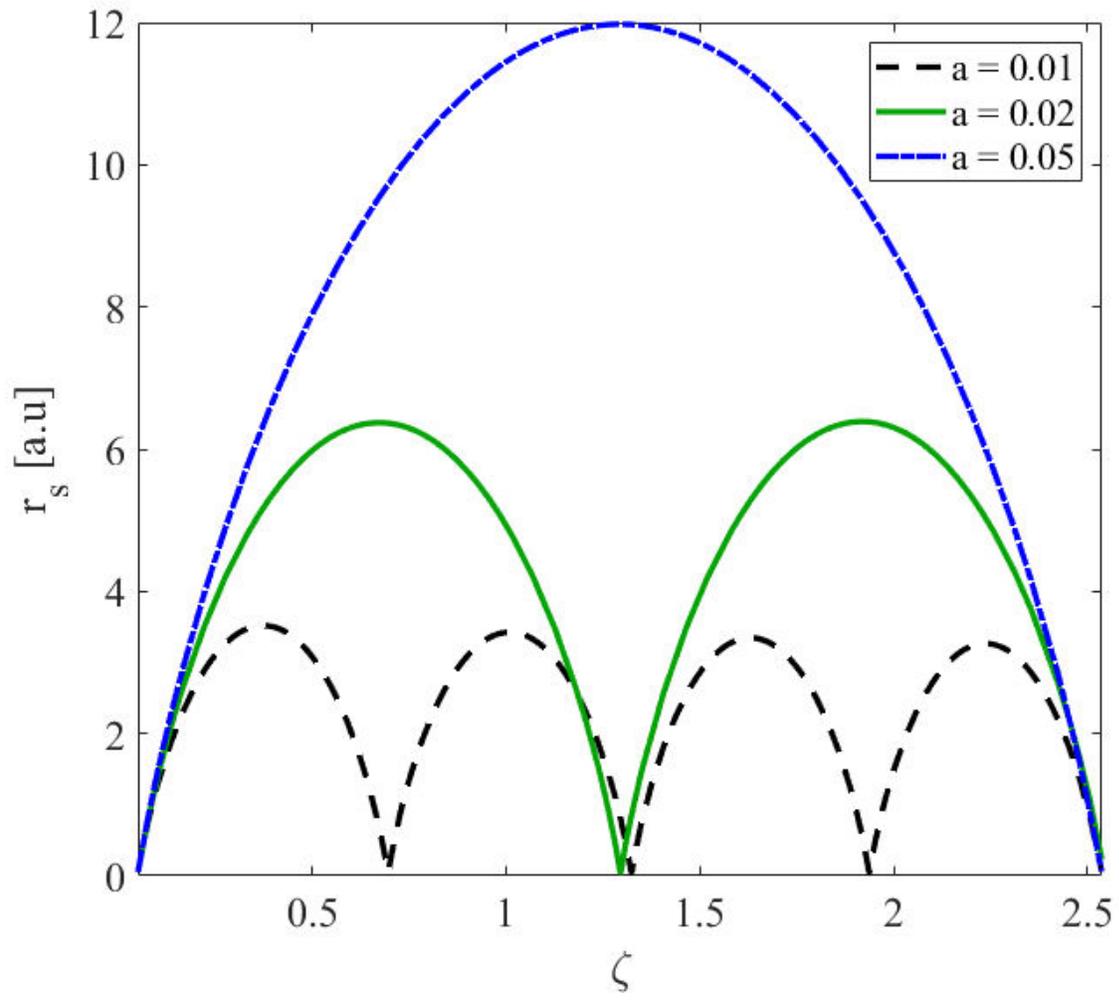

**Fig. 8.**